\documentclass[12pt,preprint]{aastex}
\def \aj {AJ}
\def \mnras {MNRAS}
\def \pasp {PASP}
\def \apj {ApJ}

\def \apjl {ApJL}
\def \aap {A\&A}
\def \nat {Nature}
\def \araa {ARAA}

\newcommand{\msun} {M$_{\odot}$}

\def\lesssim{\mathrel{\hbox{\rlap{\hbox{\lower4pt\hbox{$\sim$}}}\hbox{$<$}}}}
\def\gtrsim{\mathrel{\hbox{\rlap{\hbox{\lower4pt\hbox{$\sim$}}}\hbox{$>$}}}}

\shorttitle{The nature of SN 2010gx}
\shortauthors{Pastorello et al.}

\begin{document}

\title{Ultra-bright optical transients are linked with type Ic supernovae}

\author{A. Pastorello\altaffilmark{1}\email{a.pastorello@qub.ac.uk} S. J. Smartt\altaffilmark{1}, M. T. Botticella\altaffilmark{1}, K. Maguire\altaffilmark{1}, M. Fraser\altaffilmark{1},  K. Smith\altaffilmark{1}, R. Kotak\altaffilmark{1}, 
L. Magill\altaffilmark{1}, S. Valenti\altaffilmark{1}, D. R. Young\altaffilmark{1}, S. Gezari\altaffilmark{2,3}, F. Bresolin\altaffilmark{4}, R. Kudritzki\altaffilmark{4},D. A. Howell\altaffilmark{5}, A. Rest\altaffilmark{6}, N. Metcalfe\altaffilmark{7}, S.
Mattila\altaffilmark{1,8,9}, E. Kankare\altaffilmark{1,8,10}, K. Y. Huang\altaffilmark{11}, Y. Urata\altaffilmark{12}, W. S. Burgett\altaffilmark{4}, K. C. Chambers\altaffilmark{4}, T. Dombeck\altaffilmark{4}, H.
Flewelling\altaffilmark{4}, T. Grav\altaffilmark{2}, J. N. Heasley\altaffilmark{4}, K. W. Hodapp\altaffilmark{4}, N. Kaiser\altaffilmark{4}, G. A. Luppino\altaffilmark{4}, R. H. Lupton\altaffilmark{13}, E. A.
Magnier\altaffilmark{4}, D. G. Monet\altaffilmark{14}, J. S. Morgan\altaffilmark{4}, P. M. Onaka\altaffilmark{4}, P. A. Price\altaffilmark{4}, P. H. Rhoads\altaffilmark{4}, W. A. Siegmund\altaffilmark{4}, C. W.
Stubbs\altaffilmark{6}, W. E. Sweeney\altaffilmark{4}, J. L. Tonry\altaffilmark{4}, R. J. Wainscoat\altaffilmark{4}, M. F. Waterson\altaffilmark{4}, C. Waters\altaffilmark{4}, and
C. G. Wynn-Williams\altaffilmark{4}.}
\altaffiltext{1}{Astrophysics Research Centre, School of Mathematics and Physics, Queen's University Belfast, Belfast BT7 1NN, United Kingdom}
\altaffiltext{2}{Department of Physics and Astronomy, Johns Hopkins University, 3400 North Charles Street, Baltimore, MD 21218, USA}
\altaffiltext{3}{Hubble Fellow}
\altaffiltext{4}{Institute for Astronomy, University of Hawaii at Manoa, Honolulu, HI 96822, USA}
\altaffiltext{5}{Las Cumbres Observatory Global Telescope Network and the Department of Physics, University of California, Santa Barbara, California 93117, USA}
\altaffiltext{6}{Department of Physics, Harvard University, Cambridge,  MA 02138, USA}
\altaffiltext{7}{Department of Physics, University of Durham, South
   Road, Durham DH1 3LE, UK}
\altaffiltext{8}{Tuorla Observatory, Department of Physics and Astronomy, University of Turku, Piikki\"o, FI 21500, Finland}
\altaffiltext{9}{Stockholm Observatory, Department of Astronomy, AlbaNova University Center, SE 106 91 Stockholm, Sweden}
\altaffiltext{10}{Nordic Optical Telescope, Apartado 474, E-38700 Santa
Cruz de La Palma, Spain}
\altaffiltext{11}{Academia Sinica Institute of Astronomy and Astrophysics, Taipei 106, Taiwan}
\altaffiltext{12}{Institute of Astronomy, National Central University, Chung-Li 32054, Taiwan}
\altaffiltext{13}{Department of Astrophysical Sciences, Princeton University, Princeton, NJ 08544, USA}
\altaffiltext{14}{US Naval Observatory, Flagstaff Station,  Flagstaff,  AZ 86001, USA}

\begin{abstract}

Recent searches by unbiased, wide-field surveys have
uncovered a group of extremely luminous optical transients. The
initial discoveries of SN 2005ap by the Texas Supernova Search and 
SCP-06F6 in a deep Hubble pencil beam survey were followed by the Palomar
Transient Factory confirmation of host redshifts for other similar
transients. The transients share the common properties of 
high optical luminosities (peak magnitudes $\sim -21$ to $-23$), 
blue colors, and a lack of H or He spectral
features. The physical mechanism that produces the luminosity is 
uncertain, with suggestions ranging from jet-driven explosion to 
pulsational pair-instability. Here we report the most detailed
photometric and spectral coverage of an ultra-bright transient (SN 2010gx)
detected in the Pan-STARRS\,1 sky survey. In common with other
transients in this family, early-time spectra show a 
blue continuum, and prominent broad absorption lines of O II. 
However, about 25d after discovery,
the spectra developed type Ic supernova features, showing the characteristic 
broad  Fe II and Si II absorption lines. Detailed,  
post-maximum follow-up may show that all SN 2005ap and SCP-06F6 type 
transients are linked to supernovae Ic. This poses 
problems in understanding the physics of the explosions: 
there is no indication from late-time photometry that the luminosity is powered by 
$^{56}$Ni,  the broad lightcurves suggest very large 
ejected masses, and the  slow spectral evolution is quite different from 
typical Ic timescales. The nature of the progenitor stars and the 
origin of the luminosity are intriguing and open questions.

\end{abstract}


\section{Introduction}

The discovery of unusual optical transients is a goal of
modern surveys. Focused supernova searches (e.g. the Texas Supernova
Search) or all-sky surveys, such as the
Panoramic Survey Telescope $\&$ Rapid Response System (Pan-STARRS),
the Catalina Real-time Transient Survey (CRTS), the Palomar Transient
Factory (PTF) and Skymapper are expected to discover a large number of
new types of stellar explosions in the next years.  The
preliminary results are remarkable, and newly discovered transients
are revolutionizing our knowledge of stellar explosions.  Ultra-bright
supernovae (SNe) associated with faint and, presumably, metal-poor
host galaxies are the most spectacular recent discoveries
\citep{qui07,gez09,mil09,gal09,you10}.  

The field has moved quickly, prompted by the unusual transient \object{SCP-06F6}, discovered in the
Hubble Space Telescope Cluster Supernova Survey \citep{bar09}.  
Its lightcurve was symmetric, with a $\sim$100d rise time
in the observed frame. The
spectrum showed  broad absorption features and the transient was 
associated with no obvious host galaxy (although a weak source,
1.5'' from the transient, was marginally detected at
magnitude $z$$\sim$25.8)\footnote{Unless specified,
  magnitudes are in the AB system.}.
Without robust constraints on the absolute magnitude for this transient, 
even the discrimination between Galactic and extra-galactic origin 
was uncertain. Possible scenarios proposed by \citet{bar09} for 
\object{SCP-06F6} were an outburst of a Galactic 
C-rich white dwarf (WD), a broad absorption lines quasar or a 
micro-lensing event, but none of them was fully convincing.
Assuming that the broad features in the spectra of \object{SCP-06F6} were the  C$_2$
Swan bands, \citet{gan09} tentatively fixed the redshift 
at z=0.14,  implying an absolute peak magnitude of about -18, 
suggesting a SN-like explosion of a C-rich Wolf-Rayet (WR) star. A tidal disruption 
of a C-rich star by a black hole \citep{ros09,sok09}, a Galactic WD-asteroid merger and a type Ia SN in a dense, C-rich wind produced by 
a companion star \citep{sok09} were also proposed as alternative explanations.

A few events have recently been discovered sharing observed properties with 
\object{SCP-06F6}. Data for a total sample of 6 objects have been presented by \citet{qui10a}.
One of them was \object{SN 2005ap}, an enigmatic object originally presented in \citet{qui07}
and classified as a peculiar, over-luminous SN IIL.
Through the detection of narrow interstellar Mg II lines, \citet{qui10a} have
definitely proved that these transients are not located in the  Galaxy or in the Local Group, but are 
relatively distant objects, with redshifts between 0.26 and 1.19. 
Consequently, they are extremely luminous, with $u$-band absolute magnitudes spanning between
-22 and -23. 
On the basis of the lack of any evidence of a slope consistent with $^{56}$Co decay in the late-time lightcurve
of both \object{SCP-06F6} and \object{SN 2005ap}, \citet{qui10a}
favored either a pulsational pair-instability outburst scenario, or core-collapse SNe powered by rapidly rotating
young magnetars.

Unfortunately, follow-up observations collected so far for these transients and 
the information available for properly studying and modeling their data 
have been incomplete.  The discovery of a relatively nearby object of this class
 caught early and followed in detail, has provided us with a new opportunity to study the 
energy output and spectral evolution of one of nature's brightest explosions. 

\section{The discovery of SN 2010gx}

The CRTS team \citep{dra09} first announced the discovery of an optical transient 
\object{(CSS100313:112547-084941)} at RA=11:25:46.71 and Dec=-08:49:41.4, 
on images obtained on March 13, 2010 \citep[magnitude 18.5,][]{mah10a}. Its optical spectrum 
showed a  blue, featureless continuum, and 
the initial redshift determination (z=0.17) was
later corrected by the same authors to z=0.23 \citep{mah10b}.

On the following day \citet{qui10b} reported the independent
discovery by the  PTF survey \citep{rau09,law09}
of the same variable source (labeled as \object{PTF10cwr})
at several epochs between March 5 and 16, whilst no object was
seen on March 4.27 UT to a limiting magnitude of 20.4. Optical spectra on March 
18.27 UT showed that \object{PTF10cwr} was a luminous SN similar to the
ultra-bright \object{SN 2005ap} \citep{qui07}.  The spectrum showed
broad  features attributed to O II \citep{qui10b}. The
presence of narrow lines attributed to a host galaxy allowed them to
estimate the redshift to z=0.23. 

In the course of the Pan-STARRS1 Telescope (PS1) 3$\pi$ survey,  we 
recovered the transient (\object{PS1-1000037}, hereafter SN 2010gx)  
between March 12 and 17
showing that its luminosity was still rising \citep{pasto10}.
\citet{pasto10} noted the presence of a faint host galaxy in archive 
SDSS images (\object{SDSS J112546.72-084942.0}) with magnitudes
$g$=22.7, $g-r$=0.3, and
confirmed the \citet{qui10b}  redshift estimate  of z=0.23. 
At this redshift, the $g$-band absolute magnitude of the host galaxy is 
about -18, similar to that of the LMC.  

\section{Observations}
\subsection{Photometry} \label{ph}

\begin{figure}
\epsscale{.80}
\plotone{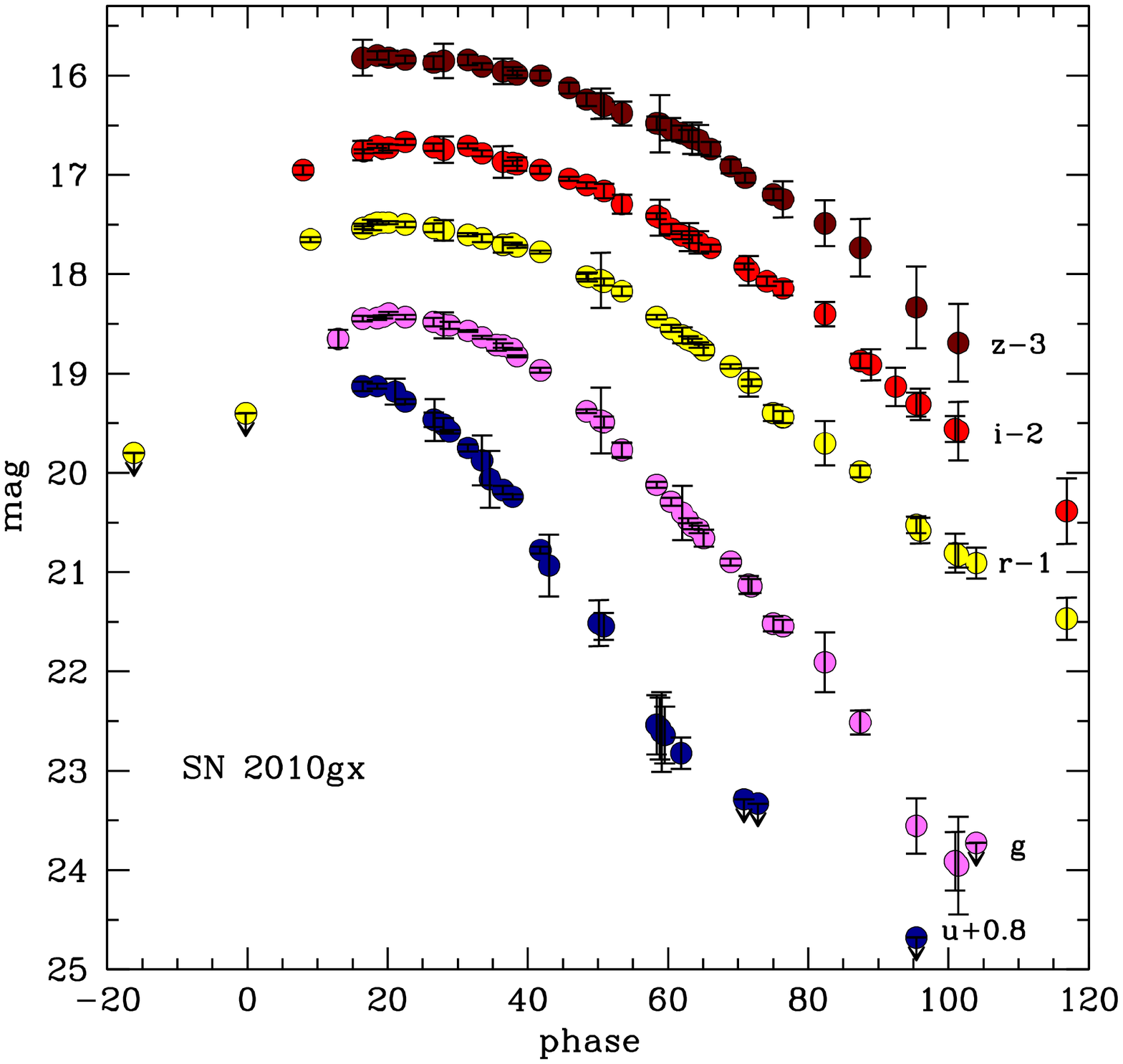}
\caption{Observed  $ugriz$ lightcurves of SN 2010gx. The phase is from JD=2455260, used as
an indicative explosion epoch. Detection limits from \protect\citet{mah10a} and
\protect\citet{qui10b} are included. 
\label{fig1}}
\end{figure}

\begin{figure}
\plotone{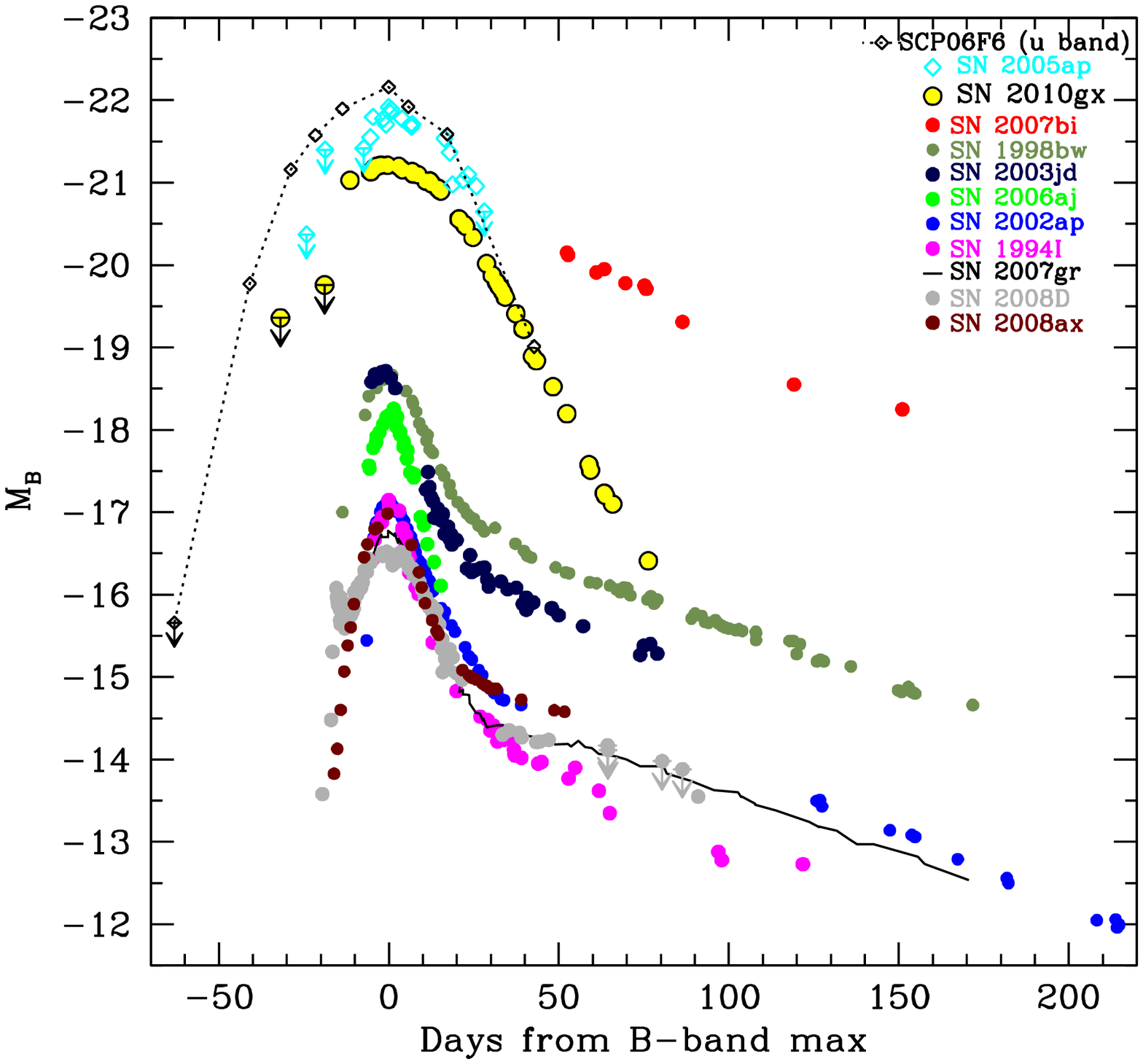}
\caption{$B$-band absolute lightcurves of SN 2010gx (Vega system) and
a number of ultra-bright events and canonical stripped-envelope SNe, including 
the type Ic SNe 1994I \protect\citep[][and references therein]{ric96}, 2002ap \protect\citep{pan02,fol03,yos03,tom06},
2006aj \protect\citep{cam06,cob06,mir06,pia06,sol06}, 2003jd \protect\citep{val08a} and 1998bw \protect\citep{gal98,mck99,sol00,pat01}; 
the type Ib SNe 2007gr \protect\citep{val08b,hun09} and 2008D \protect\citep{maz08,sod08,mod09};
the type IIb SN 2008ax \protect\citep[][]{pasto08}.
$B$-band lightcurves for the luminous SNe 2005ap, 2010gx  and 2007bi are obtained  correcting the observed broadband
photometry for time dilation and differences in effective rest-frame band (K-correction). 
The high redshift of SCP-06F6  (z=1.189) did not allow us to compute
a realistic $B$-band absolute lightcurve, so we estimated the $u$-band lightcurve (Vega system)
from the $i_{775}$-band photometry of  \protect\citet{bar09}. 
K-corrections for the luminous objects were computed using the spectra published by \protect\citet{bar09}, 
 \protect\citet{qui07}, \protect\citet{you10} and this paper.
\label{fig2}}
\end{figure}

We carried out an extensive $ugriz$ photometric follow-up campaign of \object{SN 2010gx} 
using the telescopes listed in Table \ref{tab1}.
The observed lightcurves, calibrated using 10 SDSS
stars in the field of the transient, are shown in Figure
\ref{fig1}.  The transient was discovered in
the rising phase, and its lightcurves are asymmetric. The
pre-discovery limit of March 4 \citep{qui10b} indicates that
\object{SN 2010gx} experienced a fast rise to maximum, followed
by a slower magnitude decline.  A similar asymmetry was also observed
in the lightcurve of \object{SN 2005ap} \citep{qui07}.  The
photometric evolution of \object{SN 2010gx} is somewhat different
from the bell-like shape observed in the slow-evolving lightcurve of
\object{SCP-06F6} \citep{bar09}.  
Assuming negligible host galaxy
reddening \citep[Galactic reddening of  $E(B-V)$=0.04 mag,][]{sch98} 
and accounting for redshift effects\footnote{Time dilation and 
K-correction (computed using our \object{SN 2010gx} spectra), with the latter producing Johnson $B$-band
from observed SDSS $r$-band.}, 
an absolute rest-frame peak magnitude of $M_B$$\approx$-21.2 (Vega system) is
determined for \object{SN 2010gx}. In Figure \ref{fig2} we compare the
rest-frame, $B$-band absolute lightcurve of \object{SN 2010gx} with those of a few ultra-bright events and classical type Ib/c SNe, 
including broad-lined energetic SNe Ic. 
The epoch of the $B$-band maximum for \object{SN 2010gx} was computed with a low-order
polynomial fit to the lightcurve, and found to be at JD=2455283$\pm$2.  The absolute peak magnitude of \object{SN 2010gx} is
slightly fainter than that of \object{SN 2005ap}\footnote{Note that only unfiltered photometry is available for 
SN 2005ap, calibrated using USNO-B1.0 $R2$ magnitudes \protect\citep{qui07}.}, while
no direct comparison is possible with the peculiar \object{SN 2007bi}
\citep[observed well past-maximum in the $B$-band,][]{you10} and
\object{SCP-06F6} \citep[][for which a reliable rest-frame absolute
lightcurve was computed only for the $u$-band]{bar09}.  However,
\object{SN 2010gx} appears to be 2.5-5 mag brighter 
than SNe Ib/c reported in Figure \ref{fig2}, and its overall evolution is
much slower than that of normal Ib/c events, although faster than
that experienced by \object{SN 2007bi}.

 A major difference between the lightcurves of SNe Ib/c and \object{SN 2010gx} is the {\it apparent} lack of a radioactive tail, in analogy to that
observed in the case of other objects of the Quimby et al. sample. However, with the data collected so far,  we cannot exclude the possibility 
that the lightcurve flattens onto a radioactive tail at later epochs. In that case, the expected amount of $^{56}$Ni ejected by  
\object{SN 2010gx} would be comparable (or only marginally higher, e.g. $\lesssim$1M$_\odot$) to that of type Ib/c SNe.

\subsection{Spectroscopy} \label{sp}

\begin{figure}
\plotone{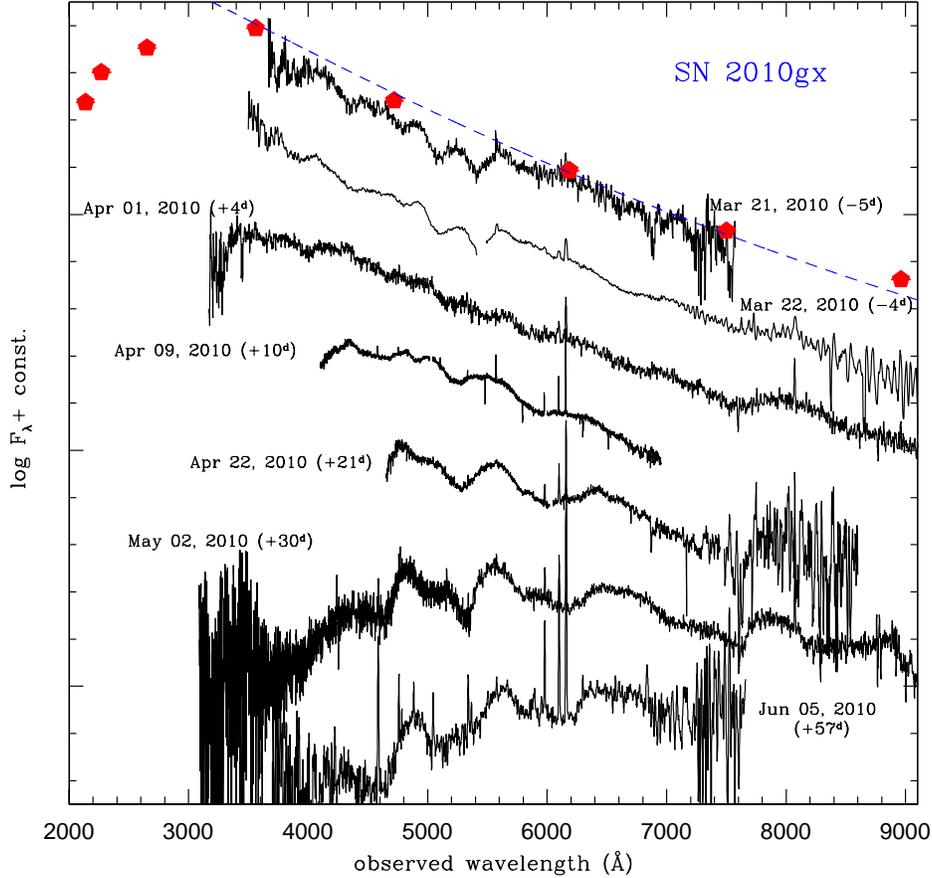}
\caption{Spectral evolution of SN 2010gx. 
All spectra are in the observed frame. The phases in parentheses are relative 
to the $B$-band maximum. The red pentagons represent the observed SED
calculated using Swift-UVOT (PI: Quimby) and Liverpool Telescope photometry 
obtained between March 19 and March 20. 
Early UVOT magnitudes (Vega system) are:  $uw2$=18.69$\pm$0.07  (JD=2455276.18), $um2$=18.21$\pm$0.08 (JD=2455274.64), $uw1$=17.71$\pm$0.06 (JD=2455274.65).
The deviation of the ultra-violet contribution from a hot blackbody continuum (T$_{bb}$=15000 K, dotted blue line) is probably due to line blanketing
in that region. 
\label{fig3}}
\end{figure}

\begin{figure}
\plotone{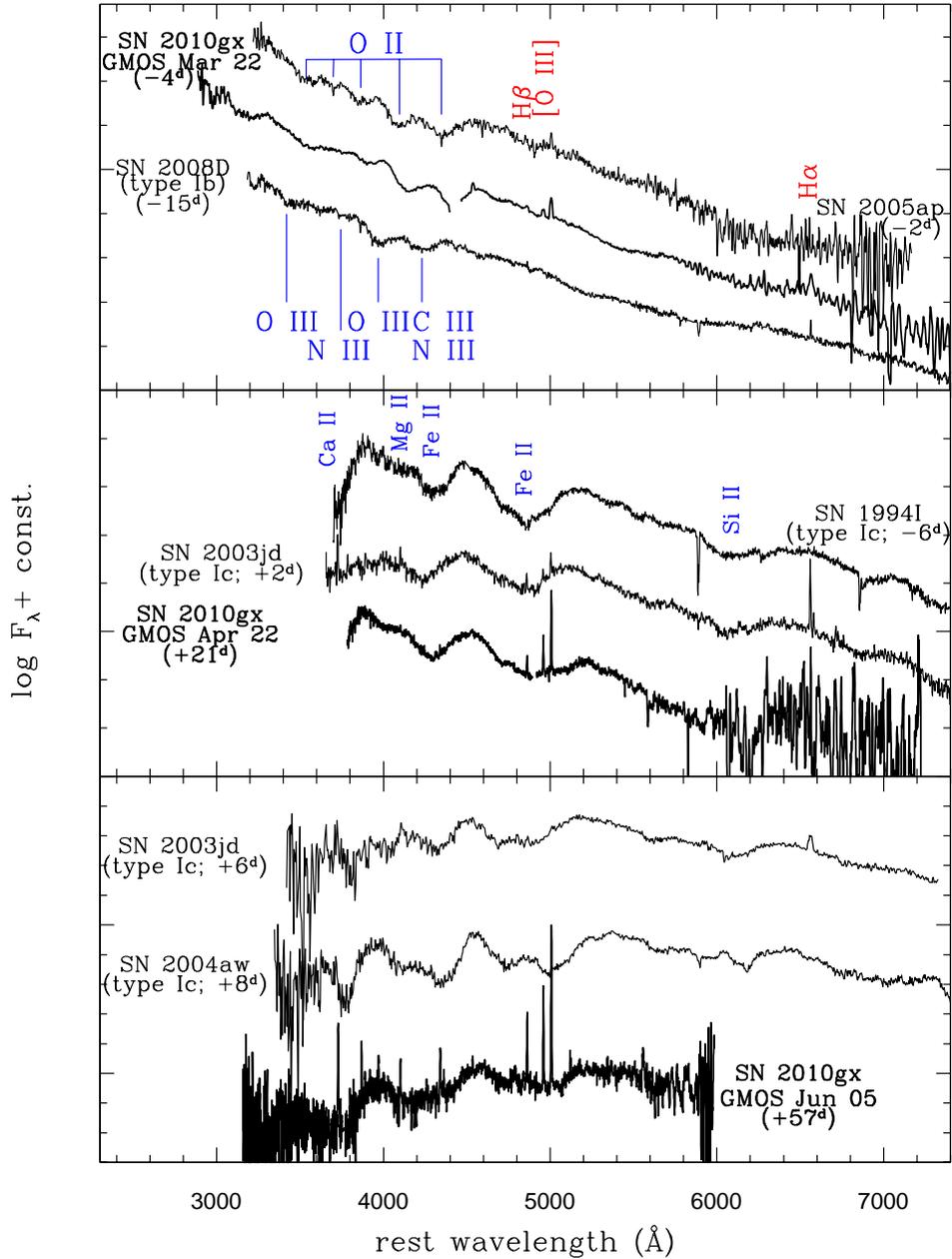}
\caption{Top: comparison of early-time spectra of SNe 2010gx and 2005ap \protect\citep{qui07} with one of 
the type Ib SN 2008D associated with the X-ray transient (XRT) 080109 \protect\citep[obtained +1.84d from XRT 080109,][]{mod09}.
All spectra show similar absorption bumps between 3500\AA~and 4500\AA, although slightly shifted in the 3 spectra. These have been tentatively identified as O II features
\protect\citep{qui10a} and blends of O III/N III/C III \protect\citep{mod09}. Middle: comparison of the April 22 spectrum of SN 2010gx
with spectra of the Ic SN 1994I \protect\citep{bar96} and the moderately broad-lined SN
2003jd \protect\citep{val08a} around maximum. Now the spectrum of SN 2010gx is dominated
by broad absorptions at about 3700\AA~(Ca II H$\&$K), 4300\AA~(Mg II, blended with Fe II), 4900\AA~(Fe II, plus possibly Mg I) and 
6100\AA~(Si II). Bottom: comparison of the June 5 spectrum of SN 2010gx with later spectra of the type Ic SNe 2003jd \protect\citep{val08a} and 
2004aw \protect\citep{tau06}. The phases labelled in figure are from the $B$-band maximum. 
\label{fig4}}
\end{figure}

A sequence of spectra of \object{SN 2010gx} was obtained with the 2.56-m Nordic Optical Telescope
and the 4.2-m William Herschel Telescope (La Palma, Canary Islands), and the 8.1-m Gemini South Telescope
(Cerro Pach\'on, Chile). Pre-maximum spectra obtained on
March 21 and 22 show a very blue continuum (with a blackbody temperature $T_{\rm bb}=15000\pm1700$ K)
with broad absorption features below $\sim$5700\AA~ (Figure \ref{fig3}). Weak, narrow emission lines
(H$\alpha$, H$\beta$ and the [O II] doublet at 4959,5007\AA) of the host galaxy  are also visible, confirming the identification of
the broad features as O II, also identified by \citet{qui10a} 
in the spectra of \object{SN 2005ap} (Figure \ref{fig4}, top). 

A spectrum obtained on April 1 (+4d) is still blue (T$_{bb}\simeq13000\pm1200$ K) but is 
almost featureless. A significant evolution of
the spectra of \object{SN 2010gx} then occurred at 10-20d after
peak. At these epochs the spectra show very broad P-Cygni absorptions of 
Ca II, Fe II and Si II, very similar to those
observed in spectra of young SNe Ic \citep{fil97}. The subsequent spectrum, obtained on May 2 (+30d), is
markedly more similar to those of SNe Ic soon after maximum
light. Finally, a further spectrum was obtained on Jun 5 (+57d), and showed only a mild evolution in  the spectral features. 

The spectral evolution of  \object{SN 2010gx}  from  an \object{SCP-06F6}-like event to a type Ic SN provides
an unexpected clue for understanding the evolutionary path of this class of transients. In order to produce O II features \citep[together with the 
Si III, C II and Mg II lines observed in spectra of other objects of this family, see][and Figure \ref{fig4}, top]{qui10b},  high photospheric temperatures are
necessary. Interestingly, \citet{mod09} noted a short-life ``W'' feature in a very early spectrum of the type Ib  \object{SN 2008D}. 
That feature, visible in a spectrum taken 1.84d after the X-ray flash \object{080109} associated with the SN, disappeared 1d later.  
\citet{mod09} noted striking similarity with the early-time spectrum of
\object{SN 2005ap} and \citep[following][]{qui07} tentatively identified such short-life features as a combination of O III, N III 
and C III lines. However, the ``W'' feature in \object{SN 2008D} is slightly blueshifted compared to the analogous feature 
visible in the early spectra of \object{SNe 2005ap} and \object{2010gx}.
Therefore, fleeting lines due to ionized intermediate-mass elements
could be  common in very early spectra of some type Ib/c SNe. 
However, these lines are visible for several weeks after the explosion in \object{SCP-06F6}-like objects, 
which is likely due to higher densities and temperatures of the  ejecta  which persist for
longer than in canonical SNe Ib/c.

As the SN expands, the ejecta become cooler and other broad lines appear (Ca II, Mg II, Fe II and Si II). These features are commonly visible in Ib/c 
spectra around maximum \citep{fil97}. In Figure \ref{fig4} (middle), a later
spectrum of \object{SN 2010gx}  (+21d) 
is compared with a pre-maximum spectrum of the normal type Ic
SN 1994I \citep{bar96} and a slightly post-maximum spectrum of the broad-line Ic SN 2003jd \citep{val08a}.
The striking similarity among these 3 spectra is a  confirmation that  \object{SN 2010gx} 
(and possibly all  \object{SCP-06F6}-like objects) should be considered
spectroscopically as SNe Ic, although with rather 
extreme photometric properties (Section \ref{ph}). The similarity with normal stripped-envelope SNe 
is even more evident in the comparison of the last spectrum of \object{SN 2010gx} (June 5) with 
spectra of the type Ic \object{SNe 2003jd} \citep{val08a}  and \object{2004aw} \citep{tau06} obtained about 1 week after their $B$-band peaks.

\section{The nature of ultra-bright events}

\object{SN 2010gx} provides important clues to understand
the nature of ultra-bright events.  Its
spectro-photometric similarities with this family is well established: 
high luminosity, slow-evolving
lightcurves, similar spectral properties and faint host
galaxies.  The spectral evolution of \object{SN 2010gx} now links
this family of transients to the more common type Ib/c SNe, and by
implication the progenitor stars.  The overall spectral evolution is
indeed similar to that of SNe Ib/c, although \object{SN 2010gx}
spectroscopically evolved on a much longer timescale. 
The observed parameters of \object{SN 2010gx} present 
several problems in interpreting the explosion. 
Its impressive luminosity at maximum
and slower evolution could simply  be interpreted as 
implying large photospheric radii ($L\sim R^{2}T^{4}$)  
and large ejecta masses ($\tau\sim(\kappa M /v)^{1/2}$;
for radiative diffusion from a sphere).  
The energy source for SNe Ib/c is 
the decay of radioactive isotopes,   but the 80d-long post-peak
 decline of \object{SN 2010gx} (Fig. 2) is too 
steep to be due to $^{56}$Co decay. 
It is plausible that a radioactive tail could be detected at later
phases, if the lightcurve flattens to the luminosity of the type Ic SN 1998bw.  
But this would imply $\lesssim$1M$_\odot$ of $^{56}$Co 
decaying into $^{56}$Fe to power the tail luminosity, and such a moderate
mass of $^{56}$Ni cannot account for the high bolometric luminosity at peak, 
which is $\sim$3-4$\times$10$^{44}$ erg s$^{-1}$. Assuming that the bolometric 
luminosity from the earliest PTF detection to our earliest multi-band observation
was constant, the energy radiated by \object{SN 2010gx} during the first 100d is $\sim$6$\times$10$^{51}$ erg.

The peak luminosity of SN 2010gx is quite similar to that of SN 2007bi, but
the lightcurve and spectral evolution are completely different
\citep{gal09,you10}.  The slow decay time, and appearance of strong
[Fe\,{\sc ii}] lines in SN 2007bi suggested a kinetic energy of few $\times$10$^{53}$ erg, very massive ejecta and  3-6M$_\odot$
of $^{56}$Ni synthesized. \citet{gal09}  postulated this 
was the explosion of a
100M$_\odot$ core  in a pair-instability SN. While this possibility
was also noted by  \citet{you10},  the  gravitational collapse of the C+O core of a
massive star (M$_{ZAMS}$ = 50-100M$_\odot$) is  a viable
mechanism  \citep[as recently calculated by][]{mor10}. Whatever the
explosion scenario, a large amount of $^{56}$Ni is necessary.
However, SN 2010gx is markedly different in its properties,
particularly the more rapid decay in its lightcurve indicates that 
the pair-instability scenario and a large $^{56}$Ni production are unlikely to be the
explanation.

The apparent lack of any evidence of lightcurve flattening to a 
radioactive tail led \citet{qui10a} to favor the {\em pulsational} pair-instability  
eruption scenario over a genuine SN explosion for
SN 2005ap, SCP-06F6 and other PTF SNe. In the pulsational pair-instability model, 
the  luminosity is generated by the collision of shells of material
material ejected at different times by the pulsations. 
Little or no $^{56}$Ni powers the lightcurve for long periods. 
The  outbursts are expected to be energetic, reaching very high peak luminosities 
and creating hot (T$_{eff}\approx$25000K), optically thick
photospheres \citep{woo07}. All of this is consistent with  the parameters
observed in ultra-bright SNe,  and our well sampled lightcurve 
of SN 2010gx is not too different from those calculated by \citet{woo07}. 
However, these models are for supergiant progenitors with
large,  extended H-rich envelopes. The energy released in the pulsations
is predicted to be 0.005-2$\times$10$^{51}$\,erg, in most cases
 below $10^{51}$\,erg. 
This is enough to eject 
the loosely bound envelope of an extended supergiant, but weather or not
this mechanism could eject a substantial part of a more compact WR star \citep{2010MNRAS.405.2113D}
remains to be calculated in detail.
Additionally, we do not see any sign of interaction between dense gas
shells in the form of narrow circumstellar lines. While the
pulsational pair-instability model is appealing as it can produce the 
high luminosity, it needs further consideration to determine if it is
physically  viable for H-free progenitor stars. 

Another possibility, also discussed in \citet{qui10a}, is that ultra-bright SNe are powered 
by the spin down of newly born magnetars \citep{kas10,woo09}.
A magnetar with a moderate magnetic field (B$\approx$10$^{14}$ Gauss) and spinning periods 
of 2-20ms can produce 
peak luminosities similar to that observed in SN 2010gx ($\sim$3-4$\times$10$^{44}$ erg s$^{-1}$). In addition,
magnetar-powered SN models do not need large $^{56}$Ni and total ejected masses \citep{kas10,woo09},
and can show Ic SN features \citep{woo09}.  
However our extensive lightcurve coverage of SN 2010gx shows a
faster decline than the model lightcurves, and our estimated photospheric
temperatures are a factor 2-4 higher than model predictions.

The observational  evidence presented in this paper links
\object{SN 2010gx}  \citep[and probably the entire family of
transients described by][]{qui10a}  with SNe Ic.  The very luminous
and broad lightcurve implies much larger ejecta masses than inferred
even for the broad-lined SNe Ic \citep[$\sim$8-15\msun,][]{val08a}. The close
similarity in the spectra implies the progenitor was a massive
WR star, but  the energy source powering the 
remarkable luminosity is uncertain. In fact we have a SN-like transient
which does not comfortably match any of the known SN scenarios, 
i.e. core-collapse and $^{56}$Ni-powered explosion, 
pair-instability, pulsational pair-instability nor magnetar-powered event.

The key diagnostics in the future will be late-time photometric monitoring after solar
conjunction, and very early detection of
new events. The presence of a late-time lightcurve tail with a slope roughly
consistent with  the $^{56}$Co decay could  support a real SN
explosion, and the early rise-time can help to determine the progenitor
radius and possibly signs of interaction between colliding shells. 
The recent suggestion that the most massive stars in the LMC may be up to 
320\msun ~\citep{crow10} could lead to more diverse SN progenitor populations
than is currently appreciated \citep{sma09}.

\acknowledgments
The PS1 Surveys have been made possible through contributions of the  
Institute for Astronomy at the University of
Hawai’i in Manoa, the Pan-STARRS Project Office, the Max-Planck  
Society and its participating institutes, the Max
Planck Institute for Astronomy, Heidelberg and the Max Planck  
Institute for Extraterrestrial Physics, Garching, The Johns
Hopkins University, the University of Durham, the University of  
Edinburgh, the Queen’s University Belfast, the Harvard-
Smithsonian Center for Astrophysics, and the Los Cumbres Observatory  
Global Telescope Network, Incorporated.
This work was conducted as part of a EURYI scheme award.

\clearpage

\clearpage

\clearpage

\appendix

\begin{table}
\begin{center}
\tiny
\caption{Observed (non K-corrected) photometry of SN 2010gx (AB mag) plus associated errors. 
UVOT-$u$ Swift, and R and I NOT data have been converted to SDSS magnitudes.
Column 2 reports the phases with respect to the $B$-band maximum. \label{tab1}}

\begin{tabular}{ccccccccc}
\tableline\tableline
Date & JD & phase$^a$ & u & g & r & i & z & telescope  \\
\tableline
12/03/2010   & 2455267.95  & -12.2&                  &                  &                 & 18.95  (0.05)   &                 &      PS1 \\
13/03/2010   & 2455268.99  & -11.4&                  &                  & 18.65  (0.03)   &                 &                 &      PS1 \\ 
17/03/2010   & 2455272.96  & -8.2 &                  &  18.65  (0.09)   &                 &                 &                 &      PS1 \\
20/03/2010   & 2455276.42  & -5.3 &   18.33  (0.05)  &  18.45  (0.03)   & 18.54  (0.05)   & 18.76  (0.10)   & 18.82  (0.18)   &      LT  \\
21/03/2010   & 2455276.61  & -5.2 &                  &                  & 18.53  (0.02)   & 18.76  (0.02)   &                 &      NOT \\
22/03/2010   & 2455277.77  & -4.3 &                  &                  & 18.50  (0.05)   &                 &                 &      GS  \\
22-23/03/2010& 2455278.50  & -3.7 &   18.33  (0.02)  &  18.44  (0.02)   & 18.48  (0.02)   & 18.71  (0.02)   & 18.80  (0.04)   &      LT  \\
23/03/2010   & 2455279.31  & -3.0 &                  &  18.43  (0.02)   & 18.48  (0.02)   & 18.73  (0.05)   &                 &      LOT \\
24/03/2010   & 2455280.14  & -2.3 &                  &  18.40  (0.02)   & 18.48  (0.02)   & 18.72  (0.03)   & 18.82  (0.07)   &      LOT \\  
25/03/2010   & 2455281.07  & -1.6 &   18.38  (0.13)  &                  &                 &                 &                 &      UVOT\\ 
27/03/2010   & 2455282.52  & -0.4 &   18.48  (0.03)  &  18.43  (0.02)   & 18.50  (0.03)   & 18.67  (0.03)   & 18.84  (0.04)   &      LT  \\
31/03/2010   & 2455286.58  &  2.9 &   18.66  (0.07)  &  18.48  (0.04)   & 18.53  (0.05)   & 18.72  (0.04)   & 18.87  (0.07)   &      LT  \\
31/03/2010   & 2455286.70  & 3.0 &   18.67  (0.21)  &                  &                 &                 &                 &      UVOT\\ 
01/04/2010   & 2455287.99  &  4.1 &   18.72  (0.07)  &  18.51  (0.13)   & 18.56  (0.10)   & 18.75  (0.13)   & 18.85  (0.17)   &      FTN \\
02/04/2010   & 2455288.85  &  4.8 &   18.78  (0.02)  &  18.52  (0.04)   &                 &                 &                 &      FTN \\
04/04/2010   & 2455291.45  &  6.9 &   18.95  (0.04)  &  18.57  (0.01)   & 18.60  (0.02)   & 18.71  (0.02)   & 18.84  (0.05)   &      LT  \\
06/04/2010   & 2455293.48  &  8.5 &   19.08  (0.25)  &  18.64  (0.02)   & 18.64  (0.04)   & 18.78  (0.03)   & 18.91  (0.03)   &      LT  \\
08/04/2010   & 2455294.59  & 9.4 &   19.26  (0.29)  &                  &                 &                 &                 &      UVOT\\ 
09/04/2010   & 2455295.53  & 10.2 &                  &  18.71  (0.06)   &                 &                 &                 &      GS  \\
09/04/2010   & 2455296.46  & 10.9 &   19.37  (0.04)  &  18.72  (0.02)   & 18.70  (0.08)   & 18.87  (0.16)   & 18.96  (0.13)   &      LT  \\
11/04/2010   & 2455297.84  & 12.1 &   19.44  (0.02)  &  18.75  (0.01)   & 18.69  (0.01)   & 18.88  (0.03)   & 18.96  (0.04)   &      FTN \\
11/04/2010   & 2455298.48  & 12.6 &                  &  18.82  (0.01)   & 18.72  (0.01)   & 18.89  (0.07)   & 18.99 (0.04)    &      LT  \\ 
15/04/2010   & 2455301.80  & 15.3 &   19.98  (0.04)  &  18.97  (0.03)   & 18.78  (0.02)   & 18.95  (0.04)   & 19.00  (0.05)   &      FTN \\
16/04/2010   & 2455303.04  & 16.3&   20.13  (0.31)  &                  &                 &                 &                 &      UVOT\\ 
19/04/2010   & 2455305.88  & 18.6 &                  &                  &                 & 19.04  (0.02)   & 19.12  (0.06)   &      FTN \\
21/04/2010   & 2455308.38  & 20.6 &                  &  19.38  (0.02)   & 19.02  (0.03)   & 19.10  (0.03)   & 19.24  (0.07)   &      LT  \\ 
22/04/2010   & 2455308.58  & 20.8 &                  &                  & 19.03  (0.05)   &                 &                 &      GS  \\
23/04/2010   & 2455310.13  & 22.1&   20.71  (0.23)  &                  &                 &                 &                 &      UVOT\\ 
23/04/2010   & 2455310.40  & 22.3 &                  &  19.48  (0.33)   & 19.06  (0.28)   &                 & 19.28  (0.15)   &      LT  \\
24/04/2010   & 2455310.84  & 22.6 &   20.75  (0.14)  &  19.49  (0.05)   & 19.08  (0.04)   &                 &                 &      FTN \\
24/04/2010   & 2455310.88  & 22.7 &                  &                  &                 & 19.16  (0.08)   & 19.30  (0.13)   &      FTN \\
26/04/2010   & 2455313.42  & 24.7 &                  &  19.77  (0.07)   &                 &                 &                 &      LT  \\ 
26/04/2010   & 2455313.44  & 24.7 &                  &  19.77  (0.08)   & 19.17  (0.05)   & 19.30  (0.10)   & 19.38  (0.12)   &      LT  \\ 
01/05/2010   & 2455318.39  & 28.8 &   21.74  (0.30)  &  20.12  (0.03)   & 19.43  (0.02)   & 19.41  (0.03)   & 19.48  (0.07)   &      LT  \\
02/05/2010   & 2455318.84  & 29.1 &   21.77  (0.31)  &                  &                 & 19.43  (0.18)   & 19.49  (0.29)   &      FTN \\
02/05/2010   & 2455319.09  & 29.3&   21.81  (0.40)  &                  &                 &                 &                 &      UVOT\\ 
03/05/2010   & 2455319.54  & 29.7 &   21.84  (0.29)  &                  &                 &                 &                 &      LT  \\
03/05/2010   & 2455320.43  & 30.4 &                  &  20.29  (0.04)   & 19.55  (0.04)   & 19.54  (0.05)   & 19.54  (0.11)   &      LT  \\
05/05/2010   & 2455321.88  & 31.6 &   22.02  (0.16)  &                  &                 &                 &                 &      FTN \\
05/05/2010   & 2455321.94  & 31.7 &                  &                  &                 & 19.61  (0.04)   & 19.58  (0.08)   &      FTS \\ 
05/05/2010   & 2455322.03  & 31.7 &                  &  20.40  (0.27)   & 19.62  (0.08)   &                 &                 &      FTS \\  
06/05/2010   & 2455322.88  & 32.4 &                  &  20.48  (0.03)   & 19.66  (0.04)   &                 &                 &      FTS \\
06/05/2010   & 2455322.94  & 32.5 &                  &                  &                 &                 & 19.61  (0.06)   &      FTS \\
06/05/2010   & 2455323.02  & 32.5 &                  &                  &                 & 19.63  (0.14)   &                 &      FTS \\
06/05/2010   & 2455323.48  & 32.9 &                  &  20.54  (0.03)   & 19.68  (0.03)   & 19.65  (0.04)   & 19.63  (0.16)   &      LT  \\
07/05/2010   & 2455324.38  & 33.6 &                  &  20.57  (0.04)   & 19.71  (0.03)   & 19.68  (0.11)   & 19.65  (0.15)   &      LT  \\
08/05/2010   & 2455325.11  & 34.2 &                  &  20.66  (0.08)   & 19.76  (0.05)   &                 &                 &      FTS \\
09/05/2010   & 2455326.09  & 35.0 &                  &                  &                 & 19.74  (0.03)   & 19.74  (0.07)   &      FTS \\
12/05/2010   & 2455328.95  & 37.4 &                  &  20.90  (0.04)   & 19.93  (0.02)   &                 & 19.91  (0.07)   &      FTS \\
14/05/2010   & 2455330.84  & 38.9 &  $>$22.49        &                  &                 &                 &                 &      FTN \\
14/05/2010   & 2455330.90  & 38.9 &                  &                  &                 & 19.92  (0.03)   &                 &      FTS \\
14/05/2010   & 2455331.01  & 39.0 &                  &                  &                 &                 & 20.03  (0.05)   &      FTS \\
14-15/05/2010& 2455331.50  & 39.4 &                  &  21.13  (0.09)   & 20.09  (0.14)   & 19.97  (0.15)   &                 &      LT  \\
15/05/2010   & 2455331.88  & 39.7 &                  &  21.14  (0.07)   & 20.09  (0.03)   &                 &                 &      FTS \\  
16/05/2010   & 2455332.82  & 40.5 &  $>$22.53        &                  &                 &                 &                 &      FTN \\  
17/05/2010   & 2455334.09  & 41.5 &                  &                  &                 & 20.07  (0.05)   &                 &      FTS \\
18/05/2010   & 2455335.02  & 42.3 &                  &  21.52  (0.08)   & 20.40  (0.08)   &                 & 20.20  (0.06)   &      FTS \\                          
19/05/2010   & 2455336.40  & 43.4 &                  &  21.54  (0.06)   & 20.44  (0.06)   & 20.14  (0.07)   & 20.25  (0.18)   &      LT  \\
25/05/2010   & 2455342.39  & 48.3 &                  &  21.91  (0.30)   & 20.70  (0.22)   & 20.40  (0.12)   & 20.49  (0.23)   &      LT  \\ 
30/05/2010   & 2455347.42  & 52.4 &                  &  22.51  (0.12)   & 20.99  (0.06)   & 20.87  (0.07)   & 20.73  (0.29)   &      LT  \\
01/06/2010   & 2455348.95  & 53.6 &                  &                  &                 & 20.91  (0.16)   &                 &      FTS \\
04/06/2010   & 2455352.47  & 56.5 &                  &                  &                 & 21.13  (0.19)   &                 &      GS  \\
08/06/2010   & 2455355.43  & 58.9 &   $>$23.88       &  23.55  (0.28)   & 21.52  (0.09)   & 21.31  (0.12)   & 21.33  (0.41)   &      WHT \\ 
08/06/2010   & 2455355.98  & 59.3 &                  &                  & 21.58  (0.13)   & 21.31  (0.16)   &                 &      FTS \\
13/06/2010   & 2455360.97  & 63.4 &                  &  23.91  (0.30)   & 21.81  (0.20)   & 21.56  (0.13)   &                 &      FTS \\
13/06/2010   & 2455361.42  & 63.8 &                  &  23.95  (0.49)   & 21.83  (0.12)   & 21.58  (0.29)   & 21.69  (0.39)   &      LT  \\   
16/06/2010   & 2455363.98  & 65.8 &                  &  $>$23.73        & 21.91  (0.16)   &                 &                 &      FTS \\ 
29/06/2010   & 2455376.88  & 76.3 &                  &                  & 22.47  (0.21)   & 22.38  (0.33)   &                 &      FTS \\           
\tableline
\end{tabular}

\tablenotetext{a}{Corrected for time dilation.}

\tablecomments{
PS1= 1.8-m Pan-STARRS1; 
GS= 8.1-m Gemini South +GMOS; 
LT= 2.0-m Liverpool Telescope +RatCam; 
NOT= 2.56-m Nordic Optical Telescope +ALFOSC; 
LOT= 1.0-m Lulin Telescope; 
UVOT= Swift +UVOT;
FTN= 2.0-m Faulkes Telescope North +MEROPE; 
FTS= 2.0-m Faulkes Telescope South +MEROPE;
WHT= 4.2-m William Herschel Telescope +ACAM.}
\end{center}
\end{table}


\end{document}